\documentclass[%
 reprint,
 amsmath,amssymb,
 aps,
]{revtex4-2}

\usepackage{graphicx}
\usepackage{dcolumn}
\usepackage{bm}
\usepackage{bbm}
\usepackage{tensor}

\begin{document}

\preprint{APS/123-QED}

\title{AdS/CFT Duality and Holographic Renormalization Group: A Review}

\author{Han Huang$^1$}%
 \email{hh697@cornell.edu}
\affiliation{%
 $^1$Sibley School of Mechanical and Aerospace Engineering, Cornell University, New York 14853, USA
 }%

\date{\today}

\begin{abstract}
In this paper we review aspects of anti de Sitter/conformal field theory (AdS/CFT) duality and the notion of holographic renormalization group (RG) flow. We start by discussing supersymmetry and construct the $\mathcal{N}=4$ super Yang-Mills theory in $d=4$ by Kaluza-Klein dimensional reduction method. Then, we study the large-$N$ limit and how it leads to the AdS/CFT duality. Using AdS/CFT, we then study the super-gravitational dual flows to the RG flows in CFT generated by marginal and relevant deformations, using the $\mathcal{N}=4$ super Yang-Mills theory as an example. Then, we prove the Zamolodchikov $C$-theorem holographically. Finally, we discuss the Wilsonian holographic renormalization.

\end{abstract}

\maketitle


\section{\label{sec:level1}Introduction}

The gauge/gravity  duality, also often referred as the holographic theory, claims a correspondence between a $d$-dimensional strongly coupled quantum field theory and a weakly coupled $d+1$-dimensional gravitational theory. Recent development suggests wide range of applications of gauge/gravity duality, ranging from high energy physics to condensed matter physics. Particularly, it suggests a new procedure of dealing with strongly coupled quantum field theories by mapping them to weakly coupled gravitational dual theories, which are potentially simpler\cite{Bire82}. 

Anti de Sitter/conformal field theory (AdS/CFT) duality\cite{Hubeny2015,ramallo2015introduction,nuastase2015introduction} can be seen as a sub-category of gauge/gravity duality with a specific symmetry requirement. One of the most studied example of AdS/CFT duality is $\mathcal{N}=4$ super Yang-Mills ($\mathcal{N}=4$ SYM) theory\cite{Rupprecht2022}, which is a $4$-dimensional CFT. $\mathcal{N}=4$ means there are $4$ supercharges, which is the maximum number of supercharges allowed for a $4$-dimensional gauge theory. This theory is special in that it is scale-invariant even quantum mechanically, while the majority of other pure gauge theories are only scale invariant classically. 

The AdS/CFT duality has been widely applied in various field of theoretical physics, including the accurate calculation of viscosity of quark-gluon plasma in quantum chormodynamics (QCD)\cite{Kisslinger20}, which studies the strong interaction of elementary particles, as well as condensed matter systems involving strongly correlated electrons, including high-temperature superconductivity and the calculation of entanglement entropy. 

In this paper, our goal is to give a pedagogical review of the holographic renormalization group (RG) flow\cite{kardar2007statistical,collins1985renormalization}. We start from $\mathcal{N}=4$ SYM theory and AdS/CFT technique as a brief review. Then, we introduce the RG flows of $\mathcal{N}=4$ SYM theory based on a UV fixed point by deforming the theory using marginal and relevant operators. Specifically, deforming the theory by revelant operators would break the conformal symmetry and generate a RG flow from the original theory in the UV fixed point. Next, we construct the holographic RG flow starting from introducing the domain wall flow, which is a candidate of gravitational dual to the quantum field theory RG flow from the UV to the IR fixed points ($i.e.$ the interpolating flow). Utilizing this foundation, we formulate the holographic proof of the Zamolodchikov $C$-theorem\cite{zamolodchikov1986irreversibility}, which states that in 2-dimensions there exists a function $C$ such that the interpolating RG flow decreases monotonically. Finally, we outline the Wilsonian holographic renormalization, with scalar field theory as an example\cite{Heemskerk2011}.

\section{A$\mathrm{d}$S/CFT Duality}
The AdS/CFT duality cliams that a strongly coupled $D=4$ gauge theory is equivalent to a gravitational theory in $D=5$ AdS spacetime. The AdS/CFT duality typically studies a $SU(N)$ gauge theory in the large-N limit, and the gauge theory is supposed to be supersymmetric. One of the simplest examples is the $\mathcal{N}=4$ SYM theory. This theory, although a seemingly unrealistic and overly-simplified theory originated from high energy physics, was actually related to condensed matter physics. For example, a $SO(6)$ spin chain can be obtained from $\mathcal{N}=4$ SYM\cite{minahan2003bethe}. The $\mathcal{N}=4$ stands for $4$ supercharges, which is the maximum number of supercharges allowed for a $D=4$ gauge theory. In this section, we first introduce supersymmetry as a unification of bosons and fermions. Then, we derive the $d=4$ $\mathcal{N}=4$ SYM theory by Kaluza-Klein dimensional reduction from $\mathcal{N}=4$ SYM theory in $D=10$. Finally, we briefly discuss the implication of the large-$N$ limit, which gives rise to the famous AdS/CFT duality.

\subsection{Supersymmetry}
In string theory\cite{Polchinski1998StringTV}, supersymmetry\cite{Nath2020} was introduced to solve the tachyon problem, which is an unstable ground state in the bosonic string theory and assumed to travel faster than light. Supersymmetry cures this problem by imposing a so-called GSO projection on the set of states that removes the tachyon. As a consequence, the fermions can give extra energies in the zero-point energy sum so that the cancellation of quantum anomalies, which utilizes the famous $1+2+3+\dots=-1/12$, is modified to yield $D=10$ in a superstring theory, instead of $D=26$. 

Supersymmetry is a more general spacetime symmetry that maps bosons to fermions. One of its major triumphs is that it solves the hierarchy problem by cancellation between fermions and bosons contribution to the electroweak scale. Supersymmetry can be viewed as an extension of the Poincar\'e symmetry, whose generators satisfy the relations

\begin{eqnarray}
\begin{aligned}
{\left[P^\mu, P^\nu\right] } & =0 \\
{\left[M^{\mu \nu}, P^\sigma\right] } & =i\left(P^\mu \eta^{\nu \sigma}-P^\nu \eta^{\mu \sigma}\right) \\
{\left[M^{\mu \nu}, M^{\rho \sigma}\right] } & =\\
i(M^{\mu \sigma} \eta^{\nu \rho}+M^{\nu \rho} & \eta^{\mu \sigma}-M^{\mu \rho} \eta^{\nu \sigma}-M^{\nu \sigma} \eta^{\mu \rho})
\end{aligned}
\label{eq:one}
\end{eqnarray}
where $P^\mu$ are generators of translations and in $D=4$, $M^{\mu\nu}$ is given by
\begin{equation}
\left(M^{\rho \sigma}\right)_\nu^\mu=i\left(\eta^{\mu \nu} \delta_\nu^\rho-\eta^{\rho \mu} \delta_\nu^\sigma\right)
\end{equation}

Supersymmetry is characterized by a $\mathbb{Z}_2$-graded Lie algebra instead of the familiar Lie algebra in most other forms of symmetries. In our context, it simply means we can have both commutators and anti-commutators, which lets us to relax the Coleman-Mandula ``no-go" theorem a little bit. It has generators in addition to the Poincar\'e generators, which are the left-handed Weyl spinor generators $Q_\alpha^A$ and the right-handed counterpart $\bar{Q}_{\dot{\alpha}}^A$, where $A=1,\dots,\mathcal{N}$. These generators are known as the supercharges. In the simplest case $\mathcal{N}=1$, the supersymmetry algebras that need to be added are $\left[ Q_\alpha, M^{\mu\nu}\right] = (\sigma^{\mu\nu})\indices{_\alpha^\beta} Q_\beta$, $\left[Q_\alpha, P^\mu\right]=\left[\bar{Q}^{\dot{\alpha}}, P^\mu\right]=0$, $\left\{Q_\alpha, Q_\beta\right\}=0$, and $\left\{Q_\alpha, \bar{Q}_{\dot{\beta}}\right\}=2\left(\sigma^\mu\right)_{\alpha \dot{\beta}} P_\mu$. The last relation tells us that the supercharges can be viewed as the square root of spacetime translations. Also, internal symmetry generators usually commute with the spinor generators, $\left[Q_\alpha, T_i\right]=0$. An exception is the $U(1)$ automorphisms of the supersymmetry algebra known as the R symmetry, namely $\left[Q_\alpha, R\right]=Q_\alpha$ and $\left[\bar{Q}_{\dot{\alpha}}, R\right]=-\bar{Q}_{\dot{\alpha}}$, where $R$ is the $U(1)$ generator. 

In the more general case, $i.e.$ the extended supersymmetry with $\mathcal{N}>1$, there are additional algebras,
\begin{eqnarray}
\left\{Q_\alpha^A, \bar{Q}_{\dot{\beta} B}\right\}=2\left(\sigma^\mu\right)_{\alpha \dot{\beta}} P_\mu \delta^A{ }_B\\
\left\{Q_\alpha^A, Q_\beta^B\right\}=\epsilon_{\alpha \beta} Z^{A B}
\end{eqnarray}
with antisymmetryc central charges $Z^{A B}=-Z^{B A}$ commuting with all the generators. 

Because $Q_\alpha^A$ is a fermionic operator that carries spin-1/2, it has the defining property of supersymmetry
\begin{equation}
Q |F\rangle=|B\rangle \text { and } \quad \mathrm{Q}| B\rangle=|F\rangle
\end{equation}
where $|F\rangle$ and $|B\rangle$ are any bosonic and fermionic states, respectively. We can also show that in any supersymmetric multiplet, the numbers of bosons and fermions must be equal. Introduce the fermionic operator defined as
\begin{equation}
(-)^F|B\rangle=|B\rangle, \quad(-)^F|F\rangle=-|F\rangle
\end{equation}
The commutation relation with $Q_\alpha$ can be easily shown by the relation
\begin{equation}
(-)^F Q_\alpha|F\rangle=(-)^F|B\rangle=|B\rangle=Q_\alpha|F\rangle=-Q_\alpha(-)^F|F\rangle
\end{equation}
hence,
\begin{equation}
\left\{(-)^F, Q_\alpha\right\}=0
\end{equation}

Tracing over $(-)^F\left\{Q_\alpha, \bar{Q}_{\dot{\beta}}\right\}$ gives 0. On the other hand, using $\left\{Q_\alpha, \bar{Q}_{\dot{\beta}}\right\}=2\left(\sigma^\mu\right)_{\alpha \dot{\beta}} P_\mu$, the trace also gives $
2\left(\sigma^\mu\right)_{\alpha \dot{\beta}} p_\mu \operatorname{Tr}\left\{(-)^F\right\}
$. Equating the two results gives 
\begin{equation}
\begin{aligned}
\operatorname{Tr} & \left\{(-)^F\right\}\\&=\sum_{\text {bosons }}\left\langle B\left|(-)^F\right| B\right\rangle+\sum_{\text {fermions }}\left\langle F\left|(-)^F\right| F\right\rangle \\
& =\sum_{\text {bosons }}\langle B \mid B\rangle-\sum_{\text {fermions }}\langle F \mid F\rangle=0.
\end{aligned}
\end{equation}
\subsection{$\mathcal{N}=4$ Super Yang-Mills Theory}
The maximum amount of supercharges in $D=4$ and with a representation of spin $\leq 1$ is $4$, corresponding to 16 presaerved Poincar\'e supercharges. More supercharges would require gravity to be present (spin = 2). The motivation of the $\mathcal{N}=4$ SYM theory follows from the scale invariance, which, in general, would be broken by the renormalization procedure. While pure gauge theories are scale-invariant classically, they are not scale-invariant after quantization. During renormalization, the coupling constants change with respect to the energy scale. For a theory that minimally couples $N_f$ Dirac spinors and $N_s$ scalars (transforming under the representation $\mathbf{R}$ of the gauge group) to the $SU(N)$ gauge field, the one-loop $\beta$-function is given by
\begin{equation}
\beta(g)=-\frac{g^3}{16 \pi^2}\left(\frac{11}{3} C(\mathbf{a d j})-\frac{4}{3} N_{\mathrm{f}} C(\mathbf{R})\right)
\end{equation}
where $g$ is the coupling constant, $C(\mathbf{R})$ is defined as $\operatorname{Tr} \left\{T_a T_b\right\} =C(\mathbf{R}) \delta_{ab}$, and $\mathbf{adj}$ denotes the adjoint representation, under which the gauge field transforms. For Dirac fermions transforming the fundamental representation of $SU(N)$, $C(\mathbf{fund})=1/2$, the $\beta$-function reads
\begin{equation}
\beta(g)=-\frac{g^3}{48 \pi^2}\left(11 N-2 N_{\mathrm{f}}\right)
\end{equation}

For $N_f<11/2 N$, the overall negative sign leads to asymptotic freedom for a general $SU(N)$ theory, which breaks the scale invariance. However, for certain special $SU(N)$ theories, the $\beta$-function can be made vanish by adding certain matter fields, including Weyl fermions and real scalar fields. In such a theory, the $\beta$-function reads
\begin{equation}
\beta\left(g_{\mathrm{YM}}\right)=-\frac{g_{\mathrm{YM}}^3}{48 \pi^2} N\left(11-2 N_f-\frac{1}{2} N_s\right)
\end{equation}
where $N_f$ now is the number of Weyl fermions, and $N_s$ is the number of real scalars. In $\mathcal{N}=4$ SYM, there are exactly 4 Weyl fermions $\lambda^a_\alpha$ ($a=1,\dots,4$), 6 real scalars $\phi^i$ ($i=1,\dots,6$), and 1 vector field $A_\mu$. 

The direct consequence of scale invariance is that, combining the scale invariance the Poincar\'e $ISO(1,3)$ symmetry together, we can obtain the conformal symmetry $SO(2,4)$, hence the $\mathcal{N}=4$ SYM is a superconformal field theory (SCFT). This property is crucial in the later discussion of AdS/CFT duality. 

To derive the Lagrangian for $\mathcal{N}=4$ SYM in $D=4$, we start from $\mathcal{N}=1$ SYM in $d=10$ and reduce the dimension from 10 to 4\cite{Rupprecht2022}. The action of $\mathcal{N}=1$ SYM in $D=10$ is given by
\begin{equation}
\mathcal{S}_{10 D}=\int \mathrm{d}^{10} x \operatorname{Tr}\left(-\frac{1}{2} F_{m n} F^{m n}+\frac{i}{2} \bar{\Psi} \Gamma^m D_m \Psi\right)
\end{equation}
where the field strength tensor is given by $F_{m n}=\partial_m A_n-\partial_n A_m+i g\left[A_m, A_n\right]$ and $\Gamma^m$ is the Dirac matrices in $d=10$.

The $D=4$ theory is obtained by separating the spacetime indices into two sectors $\mu=0,\dots,3$ and $i=1,\dots,6$ and compactify the latter sector of 6-dimensional spacetime into a 6-dimensional torun $T^6$ via the Kaluza-Klein procedure. Moreover, the gauge field $A=A_m \mathrm{d}^m$ decomposes into 
\begin{equation}
A_m=\left(A_\mu\left(x^\nu\right), \phi_i\left(x^\nu\right)\right)
\end{equation}
If we allow $\phi^i$ to transform trivially and $A^\mu$ to transform as a vector under a Lorentz transformation of $x^\mu$, the components of $F_{mn}$ can be decomposed into 
\begin{equation}
F_{\mu i}=\partial_\mu \phi_i+i g\left[A_\mu, \phi_i\right]=D_\mu \phi_i, \quad F_{i j}=i g\left[\phi_i, \phi_j\right]
\end{equation}

Similarly, we also have to reduce the kinetic term $i \bar{\Psi} \Gamma^m D_m \Psi$ of the 10-dimensional Majorana-Weyl fermion $\Psi$, which matches the fermionic and bosonic degrees of freedom, on $\mathbb{R}^{3,1} \times T^6$. Define
\begin{eqnarray}
\mathrm{P}^{+} =\frac{1}{2}\left(1+\mathrm{i} \gamma^5\right)= & \left(\begin{array}{ll}
1 & 0 \\
0 & 0
\end{array}\right) \\ 
\mathrm{P}^{-} =\frac{1}{2}\left(1-\mathrm{i} \gamma^5\right)= & \left(\begin{array}{ll}
0 & 0 \\
0 & 1
\end{array}\right)
\end{eqnarray}
the Majorana-Weyl fermion $\Psi$ can be written explicitly using Weyl spinors $\lambda^a$ as
\begin{equation}
\Psi=\left(\mathrm{P}^{+} \lambda^1, \ldots, \mathrm{P}^{+} \lambda^4, \mathrm{P}^{-} \bar{\lambda}_1, \ldots, \mathrm{P}^{-} \bar{\lambda}_4\right)^{\mathrm{T}}
\end{equation}
with $\bar{\lambda}_a=\mathrm{C} \bar{\lambda}^{A \mathrm{~T}}$ and $C$ the charge conjugate operator in $d=4$.

The Dirac matrices can be written as
\begin{eqnarray}
&& \Gamma^\mu = \mathbbm{1}_8  \otimes \gamma^\mu \\
\quad \Gamma^{a b}= && \left(\begin{array}{cc}
0 & \rho^{a b} \\
\tilde{\rho}_{ab} & 0
\end{array}\right) \otimes \mathrm{i} \gamma_5, \quad a, b=1,\dots,4
\end{eqnarray}
where $\left(\rho^{ab}\right)_{cd}=\delta_{ac} \delta_{bd}-\delta_{ad} \delta_{bc}$, and its dual is simply $\quad\left(\tilde{\rho}_{ac}\right)_{cd}=\frac{1}{2} \epsilon_{abfg}\left(\rho^{fg}\right)_{cd}=\epsilon_{abcd}$. 

If we define 
\begin{equation}
\varphi_{ab}=C_{ab}^i \varphi_i, \quad \varphi^{ab}=C^{i ab} \varphi_i=\left(\varphi_{ab}\right)^*
\end{equation}
and
\begin{equation}
\varphi_{i 4}=\frac{1}{2}\left(\varphi_i+\mathrm{i} \varphi_{i+3}\right), \quad \varphi^{a b}=\frac{1}{2} \epsilon^{abcd} \varphi_{cd}=\left(\varphi_{ab}\right)^*
\end{equation}
we can explicitly determine the Clebsch-Gordon coefficients $C_{ab}^i$ to be
\begin{equation}
\begin{array}{rlrl}
\left(C^i\right)_{j 4} & =\frac{1}{2} \delta_{ij}=\left(C^i\right)^{j 4} \\ \left(C^{i+3}\right)_{j 4} & =\frac{\mathrm{i}}{2} \delta_{ij}=-\left(C^{i+3}\right)^{j4} \\
\left(C^i\right)_{jk} & =\frac{1}{2} \epsilon_{ijk}=\left(C^i\right)^{jk}\\ \left(C^{i+3}\right)_{jk} & =-\frac{\mathrm{i}}{2} \epsilon_{ijk}=-\left(C^{i+3}\right)^{j k} .
\end{array}
\end{equation}

Hence, the Lagrangian of $\mathcal{N}=4$ SYM in $D=4$ is written as
\begin{equation}
\begin{aligned}
\mathcal{L}=\operatorname{Tr}( & -\frac{1}{2 g_{\mathrm{YM}}^2} F_{\mu \nu} F^{\mu \nu}+\frac{\vartheta}{16 \pi^2} F_{\mu \nu} \tilde{F}^{\mu \nu}-i \bar{\lambda}^a \bar{\sigma}^\mu D_\mu \lambda_a \\
& -\sum_i D_\mu \phi^i D^\mu \phi^i+g_{\mathrm{YM}} \sum_{a, b, i} C^{a b}{ }_i \lambda_a\left[\phi^i, \lambda_b\right] \\
& \left.+g_{\mathrm{YM}} \sum_{a, b, i} \bar{C}_{i a b} \bar{\lambda}^a\left[\phi^i, \bar{\lambda}^b\right]+\frac{g_{\mathrm{YM}}^2}{2} \sum_{i, j}\left[\phi^i, \phi^j\right]^2\right)
\end{aligned}
\end{equation}
where the field strength tensor $F_{\mu\nu}$ is given by $F_{\mu \nu}=\partial_\mu A_\nu-\partial_\nu A_\mu+i\left[A_\mu, A_\nu\right]$, and the covariant derivative acting on the adjoint fields is given by $D_\mu \cdot=\partial_\mu \cdot+i\left[A_\mu, \cdot\right]$.
\subsection{Large-$N$ Limit and AdS/CFT Duality}
Taking the large-$N$ limit further simplifies the theory. In the large-$N$ limit, the corresponding number of colors is large, in the sense that the gauge field is written as a $N \times N$ matrix $(A^\mu)\indices{^i_j}$. 

\begin{figure}[b]
\includegraphics[width=0.45\textwidth]{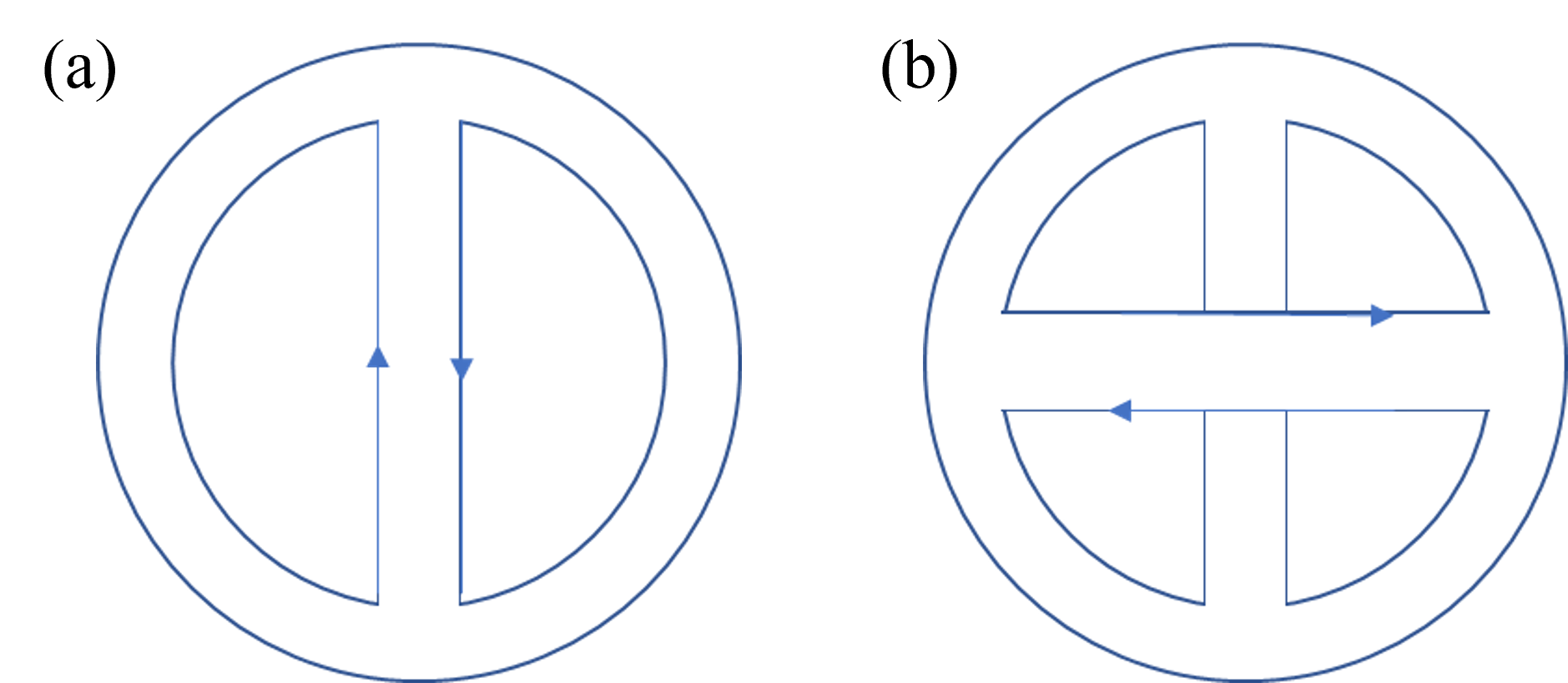}
\caption{\label{fig:1} (a) A planar diagram and (b) a non-planar diagram in the 't Hooft double-line formalism.}
\end{figure}

We now introduce the 't Hooft coupling
\begin{equation}
\lambda\equiv g_{\mathrm{YM}}^2 N
\end{equation}
Then the two independent paramenters, $N$ and $g_\mathrm{YM}$, can be replaced by $N$ and $\lambda$. In the large-$N$ limit, we set $N \rightarrow\infty$ so that $\lambda$ is kept fixed and large. The purpose of doing this is evident in that only the so-called planar diagrams dominates while summing over the Feynman diagrams. A planar diagram means the diagram can be drawn on a plane, while a non-planar diagram measn the diagram cannot be drawn on a plane, as shown in Fig.~\ref{fig:1}.

The Feymnan rules are as follows: each propagator is associated with a factor $\lambda/N$, each interaction vertex is associated with a factor $N/\lambda$, and each loop is associated with a factor $N$. If we denote the number of vertices as $V$, the number of propagators as $E$ (edge), and the number of loops as $F$ (face), we can obtain the relation
\begin{equation}
\left(\frac{N}{\lambda}\right)^V\left(\frac{\lambda}{N}\right)^E N^F=\lambda^{E-V} N^{V-E+F}
\end{equation}

Plugging in all possible values of ${V,E,F}$ ($i.e.$ all possible diagrams), we obtain
\begin{equation}
f_0(\lambda) N^2+f_1(\lambda) N^0+f_2(\lambda) \frac{1}{N^2}+\cdots 
\label{eq:3}
\end{equation}

In the large-$N$ limit, only the first term in Eq.~(\ref{eq:3}) dominates, which stands for planar diagrams with $f_0(\lambda)=1+\lambda^2+\lambda^3\dots$. 

Note that the double line diagrams can also be classfied according to their topology. For example, if we calculate the Euler characteristic given by $\chi=V-E+F$ and the genus $g=1-\chi/2$, we can see that the diagram in Fig.~\ref{fig:1}(a) corresponds to a sphere, while the diagram in Fig.~\ref{fig:1}(b) corresponds to a torus. 

The partition function is related to the summation of vacuum diagrams ($i.e.$ without external legs) as
\begin{equation}
\ln Z_{\text {gauge }}=\sum_{g=0}^{\infty} N^\chi f_g(\lambda)
\end{equation}
where $\sum_{g=0}^{\infty}$ means summing over all genus. Hence, the partition function of a large-$N$ gauge theory is given by a summation of all topologies of 2-dimensional surfaces. 

The coincidence is that the perturbative expansion of strings is also given as topologies of 2-dimensional surfaces, which intuitively originates from the world-sheet of open and closed strings. In this sense, we postulate that 
\begin{equation}
Z_{\text {gauge }}=Z_{\text {string }} 
\end{equation}
as the first clue of AdS/CFT duality. 

Another direct clue of AdS/CFT duality comes from the observation that the $AdS_5$ spacetime given by
\begin{equation}
\begin{gathered}
d s_{6}^2=-d X_0^2-d X_{5}^2+d X_1^2+\cdots+d X_{4}^2, \\
-X_0^2-X_{5}^2+X_1^2+\cdots+X_{4}^2=-L^2
\end{gathered}
\end{equation}
has two timelike directions and four spacelike direction, forming a symmetry $SO(2,4)$, which is the same as the $\mathcal{N}=4$ SYM theory. Then, the equivalence between the partition functions is replaced by
\begin{equation}
Z_{\mathrm{CFT}}=Z_{\mathrm{AdS}_5}
\label{eq:2}
\end{equation}
which is known as the GKP-Witten relation\cite{PhysRevD.82.046002}. Another way of stating this relation is, the information contained in the boundary CFT determines the bulk spactime. The $\mathcal{N}=4$ SYM theory in $D=4$ formulated above is related to the type IIB superstring theory on $AdS_5\times S^5$, which serves as the gravitational side. There are of course other more complicated examples of AdS/CFT duality. For example, the Sachdev–Ye–Kitaev (SYK) model is believed to have a gravitational dual called the Jackiw-Teitelboim dilaton gravity\cite{murugan2019one}.

The consequence of Eq.~(\ref{eq:2}) can be more easily seen if we take the saddle-point approximation on the gravity side, 
\begin{equation}
Z_{\mathrm{CFT}\left(N_c \gg \lambda \gg 1\right)}=e^{-\underline{\mathrm{S}_{\mathrm{E}}}}
\end{equation}
where $\underline{\mathrm{S}_{\mathrm{E}}}$ is the on-shell Euclidean action obtained by subsituting the classical $\mathrm{AdS}_5$ metric to the action. In this sense, a complicated field theory even in the large-$N$ limit can be easily evaluated using the gravitational theory. In addition, if there exists a blackhole in the $\mathrm{AdS}_5$ spacetime, we can even describe a finite-temperature gauge theory using the AdS blackhole, which is a thermodynamic system.

\section{Holographic Renormalization Group Flows}
In this section we rely on a rather generalized version of AdS/CFT duality, the gauge/gravity duality. First, we prove the Zamolodchikov $C$-theorem and study the gravitational duals of $\mathcal{N}=4$ SYM theory deformed by relevant and marginal operators. Then, we study the holographic formulation of RG and re-prove the $C$-theorem holographically. Finally, we examine the holographic formulation of the Wilsonian RG and show that the formulation has the same philosophy of the traditional formulation of holographic RG\cite{Heemskerk2011}.

\subsection{The $C$-Theorem}
The interpolating RG flow is generated by perturbing the field theory at a unstable UV fixed point using a relevant or marginal operator. The $C$-theorem states that in a conformal field theory, there exists a function $C(g)$ such that its value decreases monotonically with the RG flow from UV to IR fixed points, where $g = (g_1,g_2,\dots)$ is the family of coupling constants. At the fixed points, the $C$-funtion reduces to the central charge, which appears in the algebra of the symmetry group but commutes with all other symmetry operators. In a CFT, it may commute with all other operators and proportional to the conformal anomaly that breaks the conformal symmetry.

Our proof starts with a field theory action
\begin{equation}
S=\int d x L(g, \Lambda, x)
\end{equation}
where a is a UV length cutoff of the theory. RG looks for a equivalence between two theories ($i.e.$ the correlation functions of the two theories agree) related by a transformation
\begin{equation}
S(g, a, x)\quad\rightarrow S(R_t g,e^{t/2} a)
\end{equation}
where $R_t$ is a transformation in the parameter $g$-space. The RG flow is defined by the $\beta$-functions $\beta^i= a \partial g^i/\partial a$. The fixed points $g_0$ are defines as points in the $g$-space such that $\beta^i(g_0)=0$.

In a CFT, we can define the complex coordinates $(z, \bar{z})=\left(x^1+i x^2, x^1-i x^2\right)$. Thus the elements local energy-momentum tensor $T^{\mu\nu}$ can be redefined using the complex coordinate as $T\equiv T_{zz}$, $\bar{T}\equiv T_{\bar{z}\bar{z}}$, and $\Theta\equiv T_{z \bar{z}}$.

If the theory is renormalizable, we can expand the field $\theta$ by
\begin{equation}
\Theta=\beta^i(g) \Phi_i = \beta^i \frac{\partial}{\partial g^i} L(g, a, x)
\end{equation}
where we defined the scalar fields $\Phi_i = \frac{\partial}{\partial g^i} L(g, a, x)$. The conservation law $\partial^\mu T_{\mu \nu}=0$ is written in $(z,\bar{z})$ coordinates as $\partial_{\bar{z}} T+\partial_z \Theta=0$. 

In CFTs, the notion ``field" is generalized to any local expression we can write down. We define the most general
objects that are compatible with translational and rotational symmetries, scaling dimensions and
spin
\begin{equation}
\begin{gathered}
F=2 z^4\langle T(z, \bar{z}) T(0)\rangle \\
H_i=z^3 \bar{z}\left\langle T(z, \bar{z}) \Phi_i(0)\right\rangle \\
G_{i j}=z^2 \bar{z}^2\left\langle\Phi_i(z, \bar{z}) \Phi_j(0)\right\rangle
\end{gathered}
\end{equation}

The Callan-Symanzik equation implies that any correlation function $f(g,a,x)$ is invariant under the RG flow
\begin{equation}
d f / d t= 0 =\beta^i \partial_i f+a \partial_a f / 2 
\label{eq:4}
\end{equation}
as long as we are transforming in the scale larger than the UV cutoff. The $1/2$ factor in the last term of Eq.~( \ref{eq:4}) comes from the transformation $a\rightarrow e^{t/2} a$.

Using $\Phi_i = \frac{\partial}{\partial g^i} L(g, a, x)$ and $\partial_{\bar{z}} T+\partial_z \Theta=0$, we find that 
\begin{equation}
\begin{gathered}
\frac{1}{2} \beta^i \partial_i F(g, x)=3 \beta^i H_i-\beta^i \beta^k \partial_k H_i-\beta^k\left(\partial_k \beta^i\right) H_i
\end{gathered}
\end{equation}
and
\begin{equation}
\begin{gathered}
\beta^k \partial_k H_i+\left(\partial_i \beta^k\right) H_k-H_i= \\
-2 \beta^k G_{i k}+\beta^j \beta^k G_{i j}+\beta^j\left(\partial_i \beta^k\right) G_{j k}+\beta^j\left(\partial_j \beta^k\right) G_{i k}
\end{gathered}
\end{equation}

Define
\begin{equation}
C(g)=F(g)+4 \beta^i H_i=6 \beta^i \beta^j G_{i j}
\end{equation}
we have $\beta^i \partial_i c(g)=-12 \beta^i \beta^j G_{i j}$. From the positivity condition we know that $G_{ij}$ is positive definite, which implies that
\begin{equation}
\frac{d c}{d t}=\beta^i(g) \partial_i c(g) \leq 0
\end{equation}

\subsection{Deformations of $\mathcal{N}=4$ Super Yang-Mills Theory}
Adding marginal or relevant operators at the UV fixed point will generate a RG flow in the $\mathcal{N}=4$ SYM theory. Since the $\mathcal{N}=4$ SYM is automatically $\mathcal{N}=1$ symmetric, we can formulate the theory in the $\mathcal{N}=1$ superspace\cite{Leigh1995}. Marginal and relevant deformations are given by adding the superspsuperpotentials $W_m$ and $W_h$, respectively, in the form of
\begin{eqnarray}
W_h & =h^{i j k} \operatorname{Tr}\left(\Phi_i \Phi_j \Phi_k\right)\\
W_m & =m^{i j} \operatorname{Tr}\left(\Phi_i \Phi_j\right)
\end{eqnarray}
where $\Phi_i$ can be written in terms of 4 Weyl fermions $\lambda^a_\alpha$ ($a=1,\dots,4$), 6 real scalars $\phi^i$ ($i=1,\dots,6$), and 1 vector field $A_\mu$.
As an example, let us consider a superpotential
\begin{equation}
W_{\mathrm{LS}} \equiv h \operatorname{Tr}\left(\Phi_3\left[\Phi_1, \Phi_2\right]\right)+\frac{m}{2} \operatorname{Tr}\left(\Phi_3^2\right)
\label{eq:14}
\end{equation}
where the former term is marginal and the mass term is relevant. The R-symmetry, which is the symmetry transforming different supercharges, of the original $\mathcal{N}=4$ SYM theory is reduced from $SU(4)\simeq SO(6)$ (comes from Kaluza-Klein reduction from $d=10$) to $SU(2)\times U(1)$. The $\Phi_{1,2}$ are a $Su(2)$ doublet, while the $U(1)$ charges of the chiral superfields $\Phi_{1,2,3}$ are $(1/2,1/2,-1)$.

The $\beta$ function for $\mathcal{N}=1$ theories to all perturbation orders is given by the Novikov-Shifman-Vainshtein-Zakharov (NSVZ) $\beta$-function, given by
\begin{equation}
\beta(g)=-\frac{g^3}{8 \pi^2} \frac{3 C(\operatorname{\mathbf{a d j}}(G))-\sum_A C\left(\mathbf{R}_{\mathbf{A}}\right)\left(1-2 \gamma_{\mathrm{A}}\right)}{1-g^2 C(\mathbf{a d j}(G)) /\left(8 \pi^2\right)}
\end{equation}
where $\gamma_A$ is the anomalous dimension of $\Phi^A$.
Given that $G$ is $SU(N)$ and all fields transform in the adjoint representation, $C(\mathbf{R_A})=C(\mathbf{a d j}(G))=N$. The $\beta$-function is then given by
\begin{equation}
\beta(g)=-\frac{g^3 N}{4 \pi^2} \frac{\gamma_1+\gamma_2+\gamma_3}{1-3g^2/\left(8 \pi^2\right)}
\label{eq:6}
\end{equation}

Non-renormalization theorem in supersymmetric theories states that the superpotential does not receive quantum
corrections at any order in perturbation theory, leading to the $\beta$-funtions of $h$ and $m$ independent of $g$
\begin{eqnarray}
\beta_h=\gamma_1+\gamma_2+\gamma_3
\label{eq:7}
\\ \quad \beta_m=1-2 \gamma_3
\label{eq:8}
\end{eqnarray}

$SU(2)$ symmetry means $\gamma_1=\gamma_2$. Also, at a non-trivial fixed point, we require Eqs.~(\ref{eq:6}, \ref{eq:7}, \ref{eq:8}) to vanish, yielding 
\begin{equation}
\gamma_1=\gamma_2=-\frac{\gamma_3}{2}=-\frac{1}{4}
\end{equation}

The theory at the IR fixed point has a $\mathcal{N}=1$ superconformal symmetry $S U(2,2 \mid 1)$. This RG flow from a $\mathcal{N}=4$ UV fixed point to a $\mathcal{N}=1$ IR fixed point is called the Leigh-Strassler flow. 

We can calculate the conformal anomaly coefficients at UV and IR fixed points. In $D=4$, the conformal anomaly takes the form 
\begin{equation}
\left\langle T_\mu^\mu\right\rangle=\frac{c}{16 \pi^2} C^{\mu v \sigma \rho} C_{\mu v \sigma \rho}-\frac{a}{16 \pi^2} E 
\label{eq:15}
\end{equation}
where $C_{\mu v \sigma \rho}$ is the Weyl tensor defined as
\begin{equation}
C_{\mu \nu \alpha \beta} \equiv R_{\mu \nu \alpha \beta}-\left(g_{\mu[\alpha} R_{\beta] \nu}-g_{\nu[\alpha} R_{\beta] \mu}\right)+\frac{2}{6} g_{\mu[\alpha} g_{\beta] v} R
\end{equation}
and the Euler topological density $E$ is given by
\begin{equation}
E= R^{\mu \nu \sigma \rho} R_{\mu \nu \sigma \rho}-4 R^{\mu \nu} R_{\mu \nu}+R^2
\end{equation}
Note that in these equations $R^{\mu \nu \sigma \rho}$, $R^{\mu \nu}$, and $R$ are the Riemann tensor, Ricci tensor, and Ricci scalar, respectively. The conformal anomaly arises from the failure of the energy-momentum tensor to remain traceless under quantum corrections in a classically conformal field theory. In $\mathcal{N}=4$ SYM theory, 
\begin{equation}
\frac{a_{\mathrm{IR}}}{a_{\mathrm{UV}}}=\frac{c_{\mathrm{IR}}}{c_{\mathrm{UV}}}=\frac{27}{32}
\end{equation}

This result can be reproduced by the holographic dual RG flow, which will be discussed in the next section.

\subsection{Holographic Renormalization Group Flows}

Gauge/gravity duality reorganizes the RG procedure by obtaining the RG equation as a gradient flow equivalent to the supergravity equations of motion. In a supergravity theory, the spin-$2$ gravitational field acquires a superpartner of spin-$3/2$. In this section we study the domain wall (or kink) flows\cite{Freedman1999}, which interpolate between stationary points of a potential on the supergravity side. At different stationary points, the metrics become AdS metric with different AdS radius. The generalization from AdS/CFT duality to gauge/gravity duality is evident in that the regime away from the stationary points deviates from the AdS gravity. In gauge/gravity duality, however, we cannot identify the Lagrangian of the dual field theory in general.

We start from a gravitational model 
\begin{equation}
S=\int \mathrm{d}^{d+1} x \sqrt{-g}\left(\frac{R}{16 \pi G}-\frac{1}{2} \partial_m \phi \partial^m \phi-V(\phi)\right)
\label{eq:18}
\end{equation}
where $d$ is the spatial dimension. This theory describes a scalar field living in a non-trivial $d+1$ dimensional gravitational background.

The stationary points $\phi_i$ of the potential $V$ are defined as $V'(\phi_i)=0$. The equations of motion of  $g_{mn}$ and $\phi$ are given by
\begin{equation}
\frac{1}{\sqrt{-g}} \partial_m\left(\sqrt{-g} g^{m n} \partial_n \phi\right)-V^{\prime}(\phi)=0
\label{eq:12}
\end{equation}
and
\begin{equation}
R_{m n}-\frac{R}{2} g_{m n}=8 \pi G T_{m n}
\label{eq:10}
\end{equation}
where we have defined the energy-momentum tensor as $T_{mn}\equiv \left(\partial_m \phi \partial_n \phi-\frac{1}{2} g_{m n} (\partial_l \phi)^2  -g_{m n} V(\phi)\right)$.
If we study the structure of Eq.~(\ref{eq:10}), we see that at the station points where $V'(\phi_i)=\partial_m \phi=0$, we recover the AdS Einstein field equation with cosmological constant
\begin{equation}
\Lambda_i=8 \pi G V\left(\phi_i\right)=-\frac{d(d-1)}{L_i^2}
\end{equation}
where $L_i$ is the AdS radius at different stationary points. 

If we assume $\phi=\phi(r)$, we can solve the general equations of motion away from the stationary points by a metric\cite{ammon2015gauge} involving a warp factor $A(r)$, given by
\begin{equation}
\mathrm{d} s^2=e^{2 A(r)} \eta_{\mu v} \mathrm{~d} x^\mu \mathrm{d} x^\nu + \mathrm{d} r^2
\label{eq:11}
\end{equation}
known as the ``kink" ansatz. For $A(r)=r/L$, we can recover the AdS metric. With a more general linear function $A(r)$ and a constant $\phi$ near $r\rightarrow\pm\infty$ representing the boundary and the deep interior, we have a theory conjuectured to to be dual to a interpolating RG flow from UV to IR fixed points. To guarantee that in the UV at the AdS boundary, the field theory RG scale $\mu\rightarrow\infty$ for $r\rightarrow\infty$ and $\mu\rightarrow0$ for $r\rightarrow-\infty$, we can identify $r$ with $\mu$ as $\mu=\mu_0 \exp \left(\frac{r}{L}\right)$. The detail of this identification depends on the particular field theory RG scheme. 

We can calcualte the Einstein tensor with the metric Eq.~(\ref{eq:11}) 
\begin{eqnarray}
G^\mu{}_\nu  =(d-1) \delta^\mu{}_\nu\left(A^{\prime \prime}+\frac{d}{2}\left(A^{\prime}\right)^2\right)=8 \pi G T^\mu{}_\nu \\
G^r{ }_r  =\frac{d(d-1)}{2}\left(A^{\prime}\right)^2=8 \pi G T^r{ }_r
\end{eqnarray}

By subtracting $G^t{}_t-G^r{}_r$, we can obtain the second derivative of the warp factor $(A(r)$ as 
\begin{equation}
    A^{\prime \prime}=\frac{8 \pi G}{d-1}\left(T^t{ }_t-T^r{ }_r\right)=-\frac{8 \pi G}{d-1}\left(\phi^{\prime}\right)^2
\label{eq:17}
\end{equation}
which means that $A''\leq0$. If we combine the Einstein equations with Eq.~(\ref{eq:12}), we have
\begin{equation}
\begin{aligned}
\phi^{\prime \prime}+d A^{\prime} \phi^{\prime} & =\frac{\mathrm{d} V(\phi)}{\mathrm{d} \phi} \\
\left(\phi^{\prime}\right)^2-2 V(\phi) & =\frac{1}{8 \pi G} d(d-1)\left(A^{\prime}\right)^2
\label{eq:13}
\end{aligned}
\end{equation}

If we define an auxiliary superpotential $W(\phi)$ such that
\begin{equation}
V(\phi)=\frac{1}{2}\left(\frac{\mathrm{d} W}{\mathrm{~d} r}\right)^2-\frac{d}{d-1} W^2
\end{equation}
we can find the solution to Eq.~(\ref{eq:13}) by finding the solution to the first-order gradient flow equations
\begin{equation}
\sqrt{8 \pi G} \frac{\mathrm{d} \phi}{\mathrm{d} r}=\frac{\mathrm{d} W}{\mathrm{~d} \phi}, \quad A^{\prime}=-\frac{\sqrt{8 \pi G}}{d-1} W
\end{equation}
in which we look for a kink solution that interpolates between the two AdS spaces with different AdS radius $L_{UV}$ and $L_{IR}$. The kink solution serves as a candidate of the dual flow of the field theory RG flow. For the Leigh-Strassler flow in the perturbed $\mathcal{N}=4$ SYM theory, the gravitational flow called Freedman-Gubser-Pilch-Warner (FGPW) flow is conjectured to be the dual flow, which is obtained with the $\mathcal{N}=8$, $D=5$ supergravity in the gauge group $SO(6)$. Note that the resemblance to the $\mathcal{N}=4$ SYM theory $SU(4)\simeq SO(6)$.There are as many as 42 scalar fields, but we can reduce this number using symmetry arguments. If fact, truncating the $\mathcal{N}=8$ theory into $SU(2)$ singlets yields precisely the $SU(2)\times U(1)$ symmetry of the perturbed $\mathcal{N}=4$ SYM theory at the UV fixed point, given by Eq.~(\ref{eq:14}) If we consider the $SU(2)$ subgroup, the potential $V$ can be reorganized using the singlet fields $\phi$ and non-trivial $SU(2)$ representation $\chi$ as
\begin{equation}
V(\phi, \chi)=V_0(\phi)+V_2(\phi) \chi^2+\mathcal{O}\left(\chi^3\right)
\end{equation}
Hence, the stationary point $\bar{\phi}$ of $V_0$ corresponds to a stationary point $(\bar{\phi},\chi=0)$ of $V(\phi, \chi)$. The $SU(2)$ singlets $(\phi_2,\phi_3)$ can be viewed as dual to the relevant deformations in $\mathcal{N}=4$ SYM theory at UV point. Write $\rho=\operatorname{exp}{\frac{\phi_2}{\sqrt{6}}}$, the superpotential $W$ can be written as
\begin{equation}
W\left(\phi_2, \phi_3\right)=\frac{1}{4 L \rho^2}\left[\cosh \left(2 \phi_3\right)\left(\rho^6-2\right)-3 \rho^6-2\right]
\end{equation}
in which we can find the critical points. The maximum of the $W\left(\phi_2, \phi_3\right)=-3/2L$ is given at $\phi_2=\phi_3=0$. There are also saddle point solutions $\phi_2= (\ln 2 )/\sqrt{6}, \phi_3=\pm  (\ln 3) /2$, for which $W=-2^{2 / 3}/L$. The symmetry at these points are $SU(2)\times U(1)$ in addition to $SU(2)$.

We are interested in the kink flow between $\phi_2=0$ or $\phi_3=0$ and any of the two saddle points. This flow can (possibly) be viewed as the dual to the Leigh-Strassler flow due to the symmetry considerations, but we do not know the equation of motion to this flow yet. However, as another piece of evidence, we can obtain for $d=4$ 
\begin{equation}
\frac{c_{\mathrm{IR}}}{c_{\mathrm{UV}}}=\frac{W_{\mathrm{UV}}^3}{W_{\mathrm{IR}}^3}=\frac{27}{32}
\end{equation}
by plugging in $W_{\mathrm{UV}}=-3 /(2 L)$ and $W_{\mathrm{IR}}=-2^{2 / 3}/L$ obtained above. This result agrees with the field theory result for the Leigh-Strassler flow.

\subsection{Holographic $C$-Theorem}
Finally, let us reconsider the proof of the $C$-theorem. The holographic interpolating flow allows for a holographic proof of the $C$-theorem in all even dimensions, which also holds for higher dimensional field theories. 

By Eq.~(\ref{eq:15}), the conformal anomaly of the $\mathcal{N}=4$ SYM theory is given by
\begin{equation}
\left\langle T_\mu{ }^\mu\right\rangle=\frac{c}{8 \pi^2}\left(R^{\mu \nu} R_{\mu \nu}-\frac{1}{3} R^2\right), \quad c=\frac{N^2}{4}
\end{equation}
while on the gravity side
\begin{equation}
\begin{aligned}
\left\langle T_\mu^\mu\right\rangle & =\frac{L^3}{64 \pi G_5}  \left(R^{\mu \nu} R_{\mu \nu}- \frac{1}{3} R^2\right) \\ & G_5  =\frac{G_{10}}{\operatorname{Vol} \left(S^5\right)} =\frac{\pi L^3}{2 N^2}
\end{aligned}
\label{eq:16}
\end{equation}
where $G_5$ and $G_10$ are the Newtonian gravitational constants in 5 and 10 dimensions. Now, if we assume AdS/CFT is valid and consider the kink flow metric given by Eq.~(\ref{eq:11}) at the fixed points, we can replace the AdS radius L in Eq.~(\ref{eq:16}) by
\begin{equation}
L=\frac{1}{A^{\prime}(r)|_{\mathrm{FP}}}=
\end{equation}
Then, Eq.~(\ref{eq:16}) can be rewritten as
\begin{equation}
\begin{aligned}
\left\langle T_\mu{ }^\mu\right\rangle= C(r)  & \frac{1}{64 \pi} \left(R^{\mu \nu} R_{\mu \nu}-\frac{1}{3} R^2\right) \\ C(r) & =\frac{\pi}{G_5 A^{\prime}(r)^3}
\end{aligned}
\end{equation}
Eq.~(\ref{eq:17}) gives us $A^{\prime \prime} \leq 0$. Thus, 
\begin{equation}
C^{\prime}(r)=-3 \frac{1}{G_5} \frac{A^{\prime \prime}(r)}{A^{\prime}(r)^4} \geq 0
\end{equation}
proves the $C$-theorem holographically.

\subsection{Wilsonian Holographic Renormalization Group}
One important information embedded in the construction of holographic RG flow is the boundary ($r\rightarrow\infty$) of the spacetime corresponds to the UV (high energy) theory, while the interior of the bulk spacetime ($r\rightarrow -\infty$) corresponds to the IR (low energy) theory. Hence, in the presence of a UV cutoff on the field theory side, we can think of radial cutoff in the gravitational side. This motivates the holographic reorganization of Wilsonian RG\cite{Heemskerk2011,kardar2007statistical}. We illustrate this procedure using a scalar field theory.

The generating functional of a scalar field theory with a source $J(x)$ is given by
\begin{equation}
Z[J]=\int \mathcal{D} \Phi e^{\mathrm{i}S_{\mathrm{QFT}}+\mathrm{i}\int \mathrm{d}^d x J(x) \Phi(x)}
\end{equation}

In the Wilsonian approach, we perform the path integral only over $\phi(k)$ with $|k|<\Lambda$,
\begin{equation}
Z[J]=\int \mathcal{D} \Phi_{|k|<\Lambda} e^{\mathrm{i}S_{\mathrm{QFT}}^{\text {eff }}[\Phi ; \Lambda]+\mathrm{i}\int \mathrm{d}^d x J(x) \Phi(x)}
\label{eq:21}
\end{equation}
such that we can define the effective action as the $|k|>\Lambda$ portion of the integral
\begin{equation}
e^{\mathrm{i} S_{\mathrm{QFT}}^{\mathrm{eff}}[\Phi ; \Lambda]}=\int \mathcal{D} \Phi_{|k|>\Lambda} e^{\mathrm{i}S_{\mathrm{QFT}}[\Phi]}=e^{\mathrm{i}\left(I_0+I_{\Lambda}[\Phi]\right)}
\label{eq:23}
\end{equation}
where $I_0$ is the microscopic action and $I_\Lambda$ is the remaining part obtainted by integrating out modes with $|k|>\Lambda$. The corresponding effective Lagrangian in the position space is given by
\begin{equation}
\mathcal{L}_{\mathrm{QFT}}^{\mathrm{eff}}=-\frac{Z(\Lambda)}{2} \partial_\mu \Phi \partial^\mu \Phi- \frac{m^2(\Lambda)}{2} \Phi^2-\frac{g(\Lambda)}{4 !} \Phi^4+\mathcal{O}\left(\frac{1}{\Lambda^2}\right)
\end{equation}

Wilsonian RG introduces another cutoff $b\Lambda$ ($b<1$) and studies the behavior of the cutoff dependent constants $Z(\Lambda)$, $m^2(\Lambda)$, and $g(\Lambda)$ if we use $b\Lambda$ as the cutoff instead of $\Lambda$. The cutoff is lowered in this sense, so the degrees of freedom between momenta $b\lambda$ and $\Lambda$ are integrated out. The generating functional then becomes
\begin{equation}
Z[J]=\int \mathcal{D} \Phi_{|k|<b \Lambda} e^{\mathrm{i}S_{\mathrm{QFT}}^{\text {eff }}[\Phi ; b \Lambda]+\mathrm{i}\int \mathrm{d}^d x J(x) \Phi(x)}
\end{equation}
where we set $J(k)=0$ for $k>b\Lambda$. As a consequence, the modes with $b \Lambda<|k|<\Lambda$ no longer appear in the explicit part of the generating functional (or the new effective action), but the information encoded in those modes are reflected in the new $Z$, $m^2$, and $g$ factors. This is the so-called coarse-graining procedure.

If we integrate out the modes step-by-step, ($i.e.$ first integrate out $b_1 \Lambda<|k|<\Lambda$, then $b_2 \Lambda<|k|<b_1 \Lambda$ and so on), we obtain the $\beta$-function of $g(\Lambda)$ as
\begin{equation}
    \beta(g)=\frac{\mathrm{d}g(\lambda)}{\mathrm{d}\Lambda}
\end{equation}

On the other hand, on the gravitational side, we can define the radial cutoff $r=\epsilon$. This quantity is naturally related to the UV cutoff on the field theory side. The action is given by the scalar field part of Eq.~(\ref{eq:18}), $i.e.$ $S[\phi]=\int \mathrm{d}^{d+1} x \sqrt{-g}\left(-\frac{1}{2} \partial_m \phi \partial^m \phi-V(\phi)\right)$, and we define $\phi_i^\epsilon \equiv \phi_i(\epsilon)$. The partition function on the gravitational side is given by
\begin{equation}
Z^\epsilon\left[\phi^\epsilon\right] \equiv \int^{\phi \rightarrow \phi_\epsilon} \mathcal{D} \phi e^{\mathrm{i} S[\phi]}
\end{equation}
Since this functional integral is related to the field theory functional integral, it is natural to generalize it to
\begin{equation}
Z^\epsilon\left[\phi^\epsilon\right]=\int \mathcal{D} \Phi_{|k|<\Lambda} e^{\mathrm{i} I_0[\Phi]+\mathrm{i} \int \mathrm{d} x \phi^\epsilon(x) \mathcal{O}(x)}
\label{eq:22}
\end{equation}
where $\mathcal{O}$ is an operator composed of $\Phi$. Compare Eq.~(\ref{eq:21}) and Eq.~(\ref{eq:22}), we see that the bulk fields at the cutoff boundary $\phi^\epsilon$ can be interpreted as sources in an effective field theory below the cutoff $\lambda$. We can identify schematically $\Lambda \sim 1 / \epsilon$, but the detail of this correspondence depends on the detail of the field theory RG schematic, as discussed in the previous sections. 

The full bulk gravitational partition function with a source $h$ on the boundary can be written as 
\begin{equation}
\begin{aligned}
Z[h] & =\int \mathcal{D} \phi^\epsilon Z^\epsilon\left[\phi^\epsilon\right] Z_{\mathrm{UV}}^\epsilon\left[\phi^\epsilon, h\right] \\
& \equiv \int \mathcal{D} \phi^\epsilon\left[Z^\epsilon\left[\phi^\epsilon\right] \int_{\phi \rightarrow \phi_e}^{\phi \rightarrow h} \mathcal{D} \phi e^{\mathrm{i} S[\phi]}\right]
\end{aligned}
\end{equation}
where we defined the integral between the boundary and the radial cutoff $\epsilon$ as
\begin{equation}
Z_{\mathrm{UV}}^\epsilon\left[\phi^\epsilon, h\right]\equiv\int_{\phi \rightarrow \phi_\epsilon}^{\phi \rightarrow h} \mathcal{D} \phi e^{\mathrm{i} S[\phi]}
\end{equation}

Compare with Eq.~(\ref{eq:21}) and Eq.~(\ref{eq:23}), we have
\begin{equation}
e^{i I_A[\mathcal{O}]}=\int \mathcal{D} \phi^\epsilon \mathrm{e}^{\mathrm{i} \int \mathrm{d} x \phi^\epsilon(x) \mathcal{O}(x)} Z_{\mathrm{UV}}^\epsilon\left[\phi^\epsilon, h\right]
\label{eq:24}
\end{equation}
This equation holographically connects the field theory effective action with the gravitational theory. 

Within the gravitational theory, a solution of $\phi$ satisfying the boundary conditions $\phi=h$ and $\phi=\phi^\epsilon$ can be obtained and denoted as $\phi_{\star}$. The semiclassical limit of Eq.~(\ref{eq:24}) can be then obtained as the Legendre transformation relation  \begin{equation}
\begin{aligned}
I_{\Lambda}[\mathcal{O}]  = & \int \mathrm{d}^{d} x \phi_{\star}^\epsilon(x) \mathcal{O}(x)+S\left[\phi_{\star}^\epsilon \leftarrow \phi_{\star} \rightarrow h\right] \\
& \mathcal{O} =-\frac{\delta +S\left[\phi_{\star}^\epsilon \leftarrow \phi_{\star} \rightarrow h\right]}{\delta \phi_{\star}^\epsilon}
\end{aligned}
\end{equation}

With the radial direction play the role of time and using Hamiltonian mechanics, we can obtain the Legendre transformed Hamilton-Jacobi equation 
\begin{equation}
\begin{aligned}
& \partial_\epsilon I_{\Lambda}[\mathcal{O}]= \\
& \int \mathrm{d}^{d} x e^{A(r)}\left(-\frac{1}{2e^{2A(r)}} \mathcal{O}^2+\frac{1}{2}\left(\nabla_{(d)} \phi_{\star}^\epsilon\right)^2+V\left(\phi_{\star}^\epsilon\right)\right)
\end{aligned}
\label{eq:26}
\end{equation}
where $\phi_{\star}^\epsilon=\frac{\delta I_{\Lambda}[\mathcal{O}]}{\delta \mathcal{O}}$. 

We are ultimately interested in how the set of coupling constants $\lambda_i$ ($e.g.$ $Z(\Lambda)$, $m^2(\Lambda)$, $g(\Lambda)$, etc.) transforms with the radial cutoff $\epsilon$. Therefore, if we expand the action in terms of the operator $\mathcal{O}$ as $I_{\Lambda}[\mathcal{O}]=\sum_n \int \mathrm{d}^{d+1} x \sqrt{-\gamma} \lambda_n(x, \Lambda) \mathcal{O}(x)^n$, we can plug this expansion into Eq.~(\ref{eq:26}) and obtain the $\beta$-function for all $\lambda_i$. In this sense, the philosophy of the holographic formulation of Wilsonian RG is not much different from the traditional formulation discussed in the previous sections in that the RG flow can be mapped into a gravitational equation of motion.

\section{Conclusion}
In this paper we start from constructing the AdS/CFT duality from the most basic elements, $e.g.$ supersymmetry and the large-$N$ limit, and then generalize it to gauge/gravity duality by loosening several symmetry constraints, which ultimately arrives at holographic RG flow. We use the $\mathcal{N}=4$ SYM theory as a prototype to study the marginal and relevant deformations at the UV fixed points, which generates the interpolating RG flow, and relate this flow to the supergravity equations of motion, which is conjectured to be dual to the quantum field theory RG flow. Utilizing this conjectured duality, we show that the important $C$-theorem can be proved in both field theory and holographic methods. Finally, we discuss the recent advances of holographic RG in the Wilsonian RG picture\cite{Heemskerk2011}. We can see that the AdS/CFT based methods can serve as powerful tools in the RG procedure.

\nocite{*}

\bibliography{apssamp}

\end{document}